# Effect of the Jet Production on Pseudorapidity, Transverse Momentum and Transverse Mass Distributions of Charged Particles Produced in *pp*-Collisions at Tevatron Energy [*]

Ali Zaman[1 ;1)], Mais Suleymanov[1], Muhammad Ajaz[2] and Kamal Hussain Khan[1]

[1] Department of Physics, COMSATS Institute of Information Technology, Park Road, Islamabad, 44000, Pakistan
[2] Department of Physics, Abdul Wali Khan University Mardan, 23200, Pakistan

**Abstract:** We investigate the effects of jet production on the following parameters: pseudorapidity, transverse momentum and transverse mass distributions of secondary charged particles produced in pp-collisions at 1.8 TeV, using the HIJING code. These distributions are analyzed for the whole range and for selected six regions of the polar angle as a function of different number of jets. The obtained simulation results for these parameters are interpreted and discussed in connection to the increase observed in the multiplicity of secondary charged particles as a result of its multi-jet dependence and are also discussed in comparison with the experimental results from the CDF Collaboration.

## 1 Introduction

Jet quenching [1, 2] in high energy hadronic collisions at ultrarelativistic energies is considered one piece of experimental evidence for the formation of Quark-Gluon Plasma (QGP), a new state of strongly interacting matter formed under critical conditions of high density and high temperature. Jet quenching is the result of the energy loss by jets at the partonic level interaction [3]. In order to study jet production or jet suppression and its effects on the different parameters of the charged particles produced in high energy collisions, it is useful to know the jet dependence of parameters like charged particles multiplicity, pseudorapidity and transverse momenta etc. in *pp*-collisions. Here we present an investigation of the jet dependence of the pseudorapidity, transverse momentum and transverse mass distributions of charged particles produced in *pp*-interactions at 1.8 TeV energy using the Heavy Ion Jet Interaction Generator (HIJING) Monte Carlo Model [4, 5]. The HIJING Monte Carlo model is basically designed to simulate multiple jets and particle production in hadron-hadron (*pp*), hadron-nucleus (*pA*) and nucleus-nucleus (*AA*) interactions at collider energies. There are some special parameters in HIJING, like energy, frame (lab or centre of mass), types of the colliding nuclei, their impact parameter and some other parameters regarding the production of jets and other particles, which users have to specify or change. In particular, HIJING reproduces many inclusive spectra two particle correlations, and can explain the observed flavor and multiplicity dependence of the average transverse momentum. The concept of jets and their association with hard parton scattering has been well established in hadronic interactions and they have been proven to play a major role in every aspect of *pp* collisions at high (SPS and Tevatron) energies. At hadron-hadron level interaction, HIJING also made an important effort to address the interplay between low transverse momentum ($p_T$) non-perturbative physics and the hard perturbative Quantum Chromodynamics (pQCD) processes. This Monte Carlo model has been tested extensively against data on *pp* interaction over a wide range of collider energy. More details about HIJING can be found in References [4, 5].

This study is a continuation and confirmation of the HIJING code results on the influence of the jets on charged particles multiplicities ($N_{ch}$) in *pp*-collisions at 1.8 TeV [6], where we presented and discussed the increase observed in $N_{ch}$ as a function of different number of jets for charged particles distributions in full phase space (whole polar angle rage) and for six different regions of the polar angle $\theta$ (R1: $\theta$=0-2°, R2: $\theta$=2-4°, R3: $\theta$=4-6°, R4: $\theta$=6-10°, R5: $\theta$=10-30° and R6: $\theta$=30-90°). The reason for selection of these angular intervals R1-R6 was the statistical reliability of the data. We chose this selection of the polar angle regions from analyses of the angular distributions of the secondary charged particles. The results of distributions of $N_{ch}$ for zero and multi-jet events were found to be inconsistent with the experimental multiplicity distributions (MDs) of charged particles interpreted by fitting with the Pascal (Negative Binomial) Distributions [7].

---
[*] Supported by Higher Education Commission (HEC) Government of Pakistan under Indigenous 5000 PhD Scholarship Program Batch-IV
1) Email: ali.zaman@cern.ch

In the present work we present a study of the effects of jet production on pseudorapidity ($\eta$), transverse momentum ($p_T$) and transverse mass ($m_T$) distributions of secondary charged particles for the whole polar angle range and also for six selected regions of polar angle R1 to R6, where R1 to R6 are defined as above. The relation that defines the pseudorapidity is $\eta = -\ln\left[\tan\left(\frac{\theta}{2}\right)\right]$, where $\theta$ is the polar angle with the beam axis. The transverse momentum is defined by $p_T = \sqrt{p_x^2 + p_y^2}$ or $p_T = p\sin\theta$. The total momentum is given by $p^2 = p_x^2 + p_y^2 + p_z^2$ with $p_z$ the longitudinal momentum parallel to the beam ($z$-) axis. The transverse mass is defined by the relation $m_T = \sqrt{m^2 + p_T^2}$. Here, we have studied the results for pseudorapidity, transverse momentum and transverse mass distributions of secondary charged particles for the whole polar angle range and for six regions in comparison with the reported results for change (increase) in multiplicity of $N_{ch}$ in multi-jet events [6]. Moreover, these results are discussed in connection with the experimental results for pseudorapidity distributions of charged particles from the Collider Detector at Fermilab (CDF Collaboration) [8].

## 2  Simulation Results for *pp* Interactions at 1.8 TeV

We have analyzed $\eta$-, $p_T$- and $m_T$-distributions of the secondary charged particles produced in *pp*-collisions at 1.8 TeV centre of mass energy. The Dubna version of the HIJING code (modified by Uzhinsky [9]) is used for simulation of 100,000 events. These distributions of secondary charged particles, including protons, charged $\pi$ ($\pi^\pm$) and charged $K$ – mesons ($K^\pm$-meson) are considered. These distributions are considered for different numbers of jets ($N_{jet}$) for the whole range of the polar angle as well as in its different regions. The Number of jets was $N_{jet}$= D, i.e. the default value of jets (maximum number) taken from the HIJING Model, 0, 1 and 2 (and labeled as D, 0, 1 and 2 respectively). (In the HIJING Model, the parameter for number of jets has default values D which is the maximum number of jet production; it can be turned off i.e. 0 jets and can be fixed to any number 1, 2, and so on. For the particular case considered here since the distributions of $\eta$, $p_T$ and $m_T$ for simulation results shown below for $N_{jet}$=D are same as for $N_{jet}$=1 so we can say that default value of $N_{jet}$ is 1.) Different angular regions were selected for six ranges of the polar angle theta $\theta$ (in degrees) from R1 to R6.

### 2.1 Pseudorapidity Distributions

The full phase space $\eta$-distribution of charged particles produced in *pp*-collisions at 1.8 TeV, as a function of number of jets $N_{jet}$=D, 0, 1 and 2, superimposed with the $\eta$-distribution for $|\eta|\leq3.5$ from CDF [8] at 1.8 TeV, is shown in Fig. 1. This figure shows some plateaus in the central pseudorapidity regions. The existence of plateaus in rapidity or pseudorapidity is very important for theoretical estimation. For example, J. D. Bjorken described the space-time evolution of the hadronic matter produced in the central rapidity region in extreme nucleus-nucleus collisions [10]. He found that quark-gluon plasma is produced at a temperature of the order of 200 MeV, which was in agreement with previous measurements [11]. The author also commented that the description relies on the existence of a flat central plateau and on the applicability of hydrodynamics. Here in Fig. 1, one can see that with increasing number of jets:

- the width of the distribution/plateaus is decreased,

- the pseudorapidity density in the central area is increased.

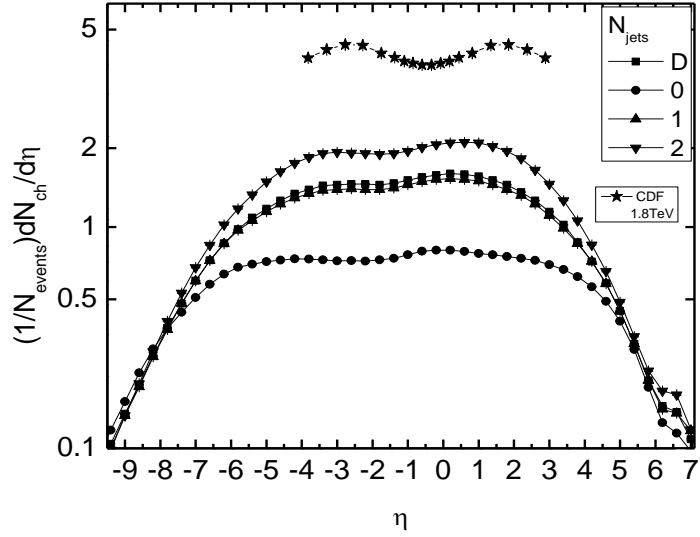

Fig. 1.  $\eta$-distribution of charged particles for $N_{jet}$=D, 0, 1 and 2 for the whole range of polar angle, superimposed with the $\eta$-distribution for $|\eta|\leq 3.5$ from CDF [8] at 1.8 TeV.

In Fig. 2 the data are presented for the corrected $\eta$ distributions (corrected for geometric and kinematic acceptance, tracking efficiency etc.) of charged particles produced in proton-antiproton collisions at $\sqrt{s}$=1800 and 630 GeV from the Collider Detector at Fermilab [8]. A measurement from the CERN SPS collider performed by UA5 at $\sqrt{s}$=546 GeV is also shown in this figure. A qualitative comparison of these data with those from Fig. 1 for the pseudorapidity region $|\eta|\leq 3.5$ demonstrates that as the $\eta$ spectra for CDF data are much higher than those for the HIJING ($N_{jet}$=2) data, this increase in pseudorapidity density is due to multi-jet events.

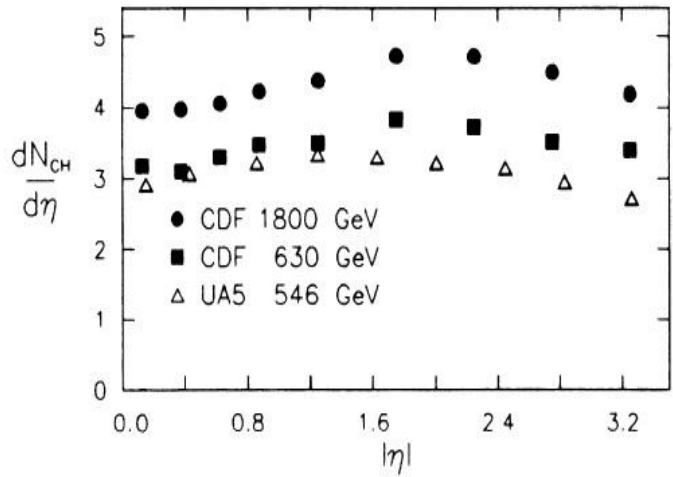

Fig. 2.  The pseudorapidity density measured by the CDF Collaboration at 1800 and 630 GeV, and by the UA5 Collaboration at 546 GeV [8].

Figure 3 (*a*) to (*f*) demonstrates the $\eta$-distribution of secondary charged particles produced in *pp*-collisions at 1.8 TeV as a function of $N_{jet}$=D, 0, 1 and 2, for the six regions of polar angle R1 to R6 respectively. One can observe that:

- for particles with polar angle less than 2 degrees (R1) the $\eta$-distributions do not depend on $N_{jet}$. However there are observed two regions (peaks) for the $\eta$-distribution of the particles separated by a sharp drop at $\eta$=8; especially for the case $N_{jet}$=2, there is a very sharp transition from the first region to the second. This effect is connected with kinematics *($\theta$-$\eta$ correlation)* because the probability of contribution to the second area ($\eta$>8) is very low due to the $\theta$-$\eta$ correlation (particles with $\eta$>8 have $\theta$~0, i.e. they are mainly leading particles, particles with energies very close to the projectile energies).

- for particles with the polar angle greater than 2 degrees, the width and height of the distributions increase with $N_{jet}$.

It is also observed in this figure that the pseudorapidity spectra for $N_{jet}$=2 are systematically higher than those for $N_{jet}$=0 for all the regions R1-R6, which shows that 2 (multi)-jet events lead to an increase in the pseudorapidity density.

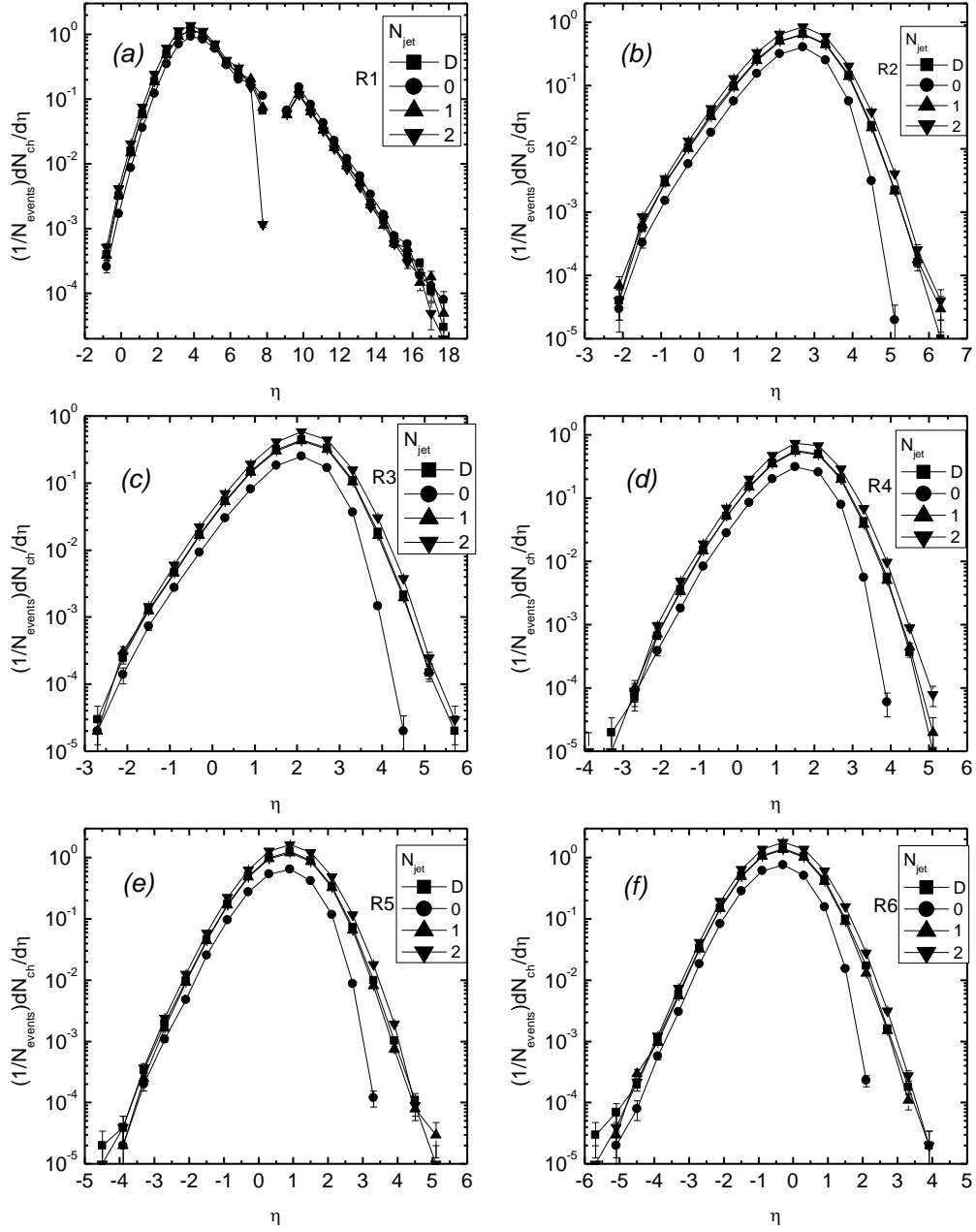

Fig. 3. $\eta$-distributions of charged particles for $N_{jet}$=D, 0, 1 and 2 in the angular regions: *(a)* R1 $\theta$=0-2° *(b)* R2 $\theta$=2-4° *(c)* R3 $\theta$=4-6° *(d)* R4 $\theta$=6-10° *(e)* R5 $\theta$=10-30° *(f)* R6 $\theta$=30-90°.

## 2.2 Transverse Momentum Distributions

In Figure 4 and Figure 5 ((a) to (f)), the transverse momentum $p_T$ distributions of secondary charged particles produced in *pp*-interactions at 1.8 TeV as a function of number of jets $N_{jet}$=D, 0, 1 and 2, are presented, for the whole range and for the six selected regions (R1 to R6) of the polar angle. We can see that with increasing the number of jets from 0 to 1, 2 or with the default values of jets the transverse momentum is increased as the jets or multi-jet events contain high $p_T$ particles.

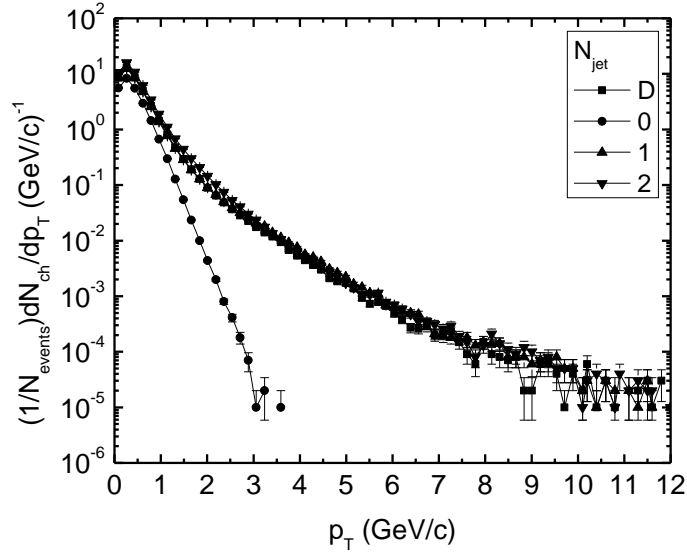

Fig. 4. $p_T$ distribution of charged particles for $N_{jet}$=D, 0, 1 and 2 for the whole range of polar angle.

The shape of the distributions in the case of $N_{jet} \geq 1$ is different in two areas of $p_T$: $p_T$<2 GeV/c and $p_T$>2 GeV/c. The slopes of the distributions in the first area are very close to that of the events with $N_{jet}$=0 (slopes of the $p_T$ distributions for the whole *θ*-range and for regions R1-R6 are shown in Fig. 6). So we could say that particles with $p_T$<2 GeV/c are produced by the same dynamics in events with different $N_{jet}$, i.e. there is no jet dependence in the $p_T$<2 GeV/c region (or zero-jet events have no contribution to the $p_T$>2 GeV/c region). The particles with $p_T$>2 GeV/c are those produced by some special dynamics, different from the particles produced with $p_T$<2 GeV/c; namely, the jet dynamics (production and hadronization of the jets) from the HIJING code. In our previous study [6] we also observed two regions for the multiplicity distribution of charged particles. We concluded that the high multiplicity regions in the $N_{ch}$-distribution correspond to the multi-jet events. The results were in good agreement with those discussed by Ugoccioni and Giovannini [7].

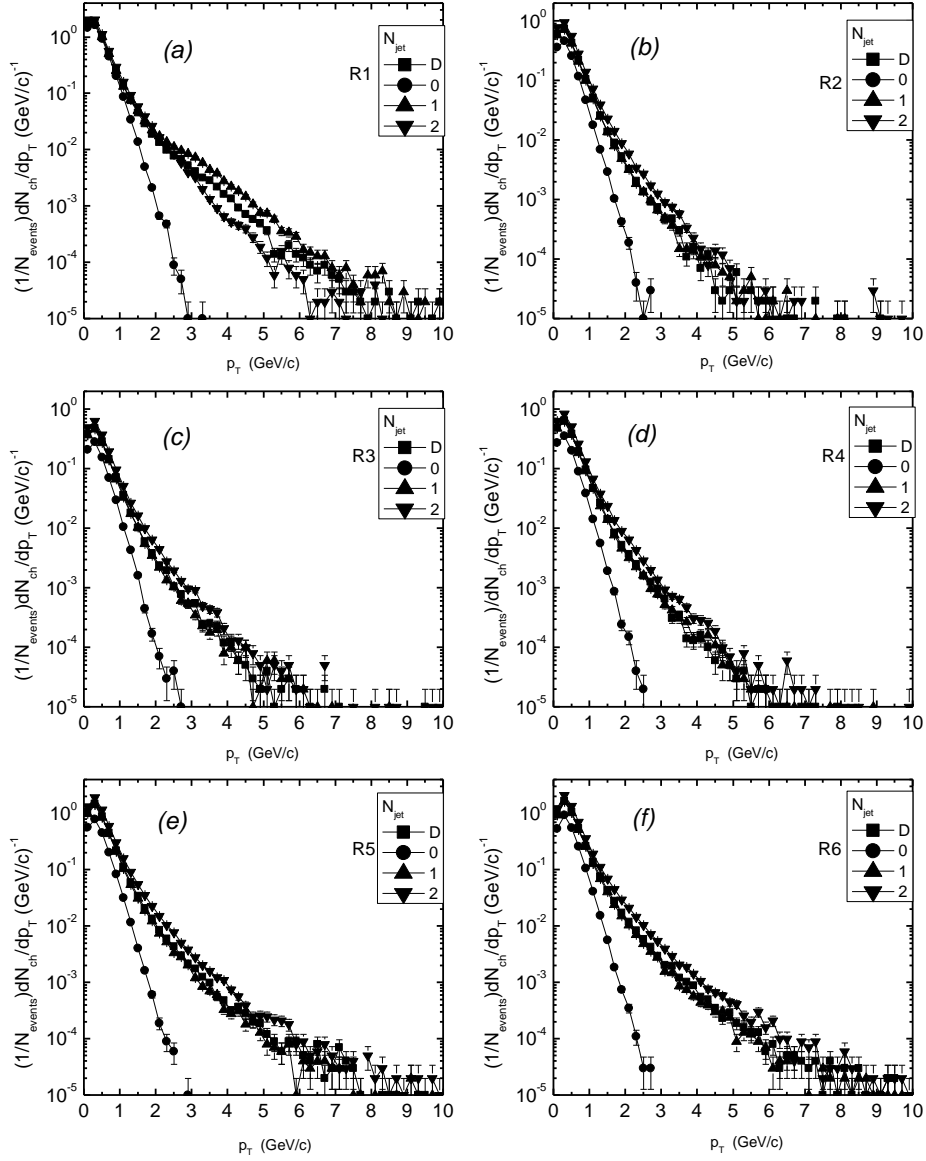

Fig. 5. $p_T$ distributions of charged particles for $N_{jet}$=D, 0, 1 and 2 in the angular regions: *(a)* R1 θ=0-2° *(b)* R2 θ=2-4° *(c)* R3 θ=4-6° *(d)* R4 θ=6-10° *(e)* R5 θ=10-30° *(f)* R6 θ=30-90°.

Figure 5 (*(a) to (f)*) demonstrates how the increase in values of the $p_T$ spectra depends on the polar angle of the produced charged particles (as $p_T = p\sin\theta$). This shows the dependence of the variation in the $p_T$ spectra with respect to the polar angle (the angular regions R1-R6) and its dependence on $N_{jets}$.

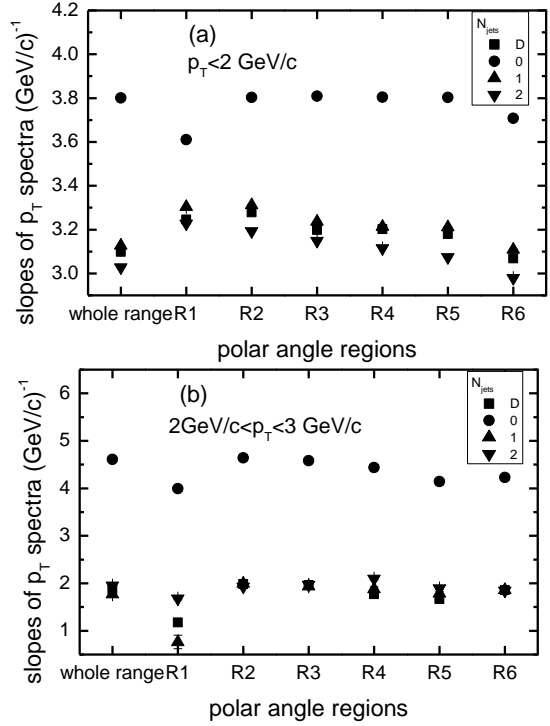

Fig. 6. Slopes of the $p_T$ spectra for the whole range and selected regions R1-R6 of the polar angle, *(a)* for $p_T<2$GeV/c and *(b)* for $2$GeV/c$<p_T<3$GeV/c

In Figure 6 (*(a)* and *(b)*), plots are shown for the slopes of the $p_T$ spectra for the whole range and selected six regions R1-R6 of the polar angle for $p_T<2$GeV/c and for $2$GeV/c$<p_T<3$GeV/c as a function of $N_{jets}$. This supports the analysis described above for the $p_T$ spectra.

## 2.3 Transverse Mass Distributions

We now consider the transverse mass $m_T$ distributions of secondary charged particles produced in *pp*-interactions at 1.8 TeV as a function of number of jets $N_{jet}$=D, 0, 1 and 2, for the whole range and the six selected regions (R1 to R6) of the polar angle. It is well known that the $m_T$ distributions are more sensitive to the temperature of the systems. Figure 7 shows the $m_T$ distributions of secondary charged particles produced in *pp*-interactions at 1.8 TeV for $N_{jet}$=D, 0, 1 and 2 for the whole polar angle range. With increasing the number of jets from 0 to 1, 2 or the default value of jets, an increase in the transverse mass is observed and here the increase in the case of multi-jet events is more than for the transverse momentum distributions as transverse mass is $m_T = \sqrt{m^2 + p_T^2}$ .

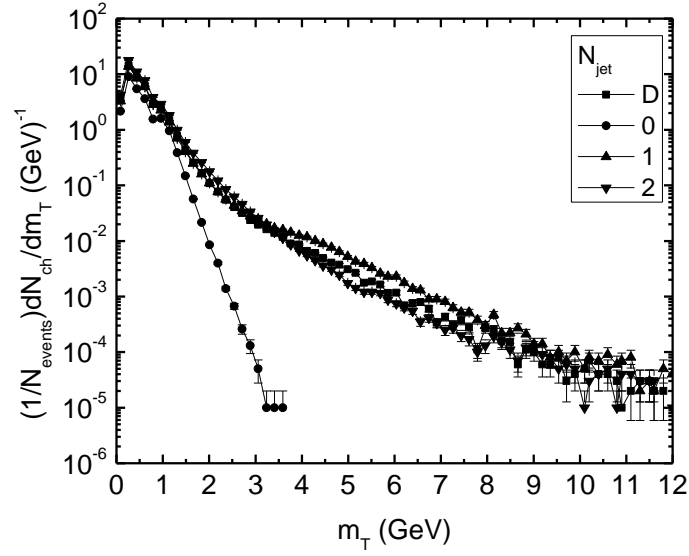

Fig. 7. $m_T$ distribution of charged particles for $N_{jet}$=D, 0, 1 and 2 for the whole range of polar angle.

Transverse mass distributions of secondary charged particles for *pp*-collisions at 1.8 TeV, for the six selected regions (R1-R6) of polar angle, are shown in Fig. 8 *(a)* to *(f)* respectively. An increase in $m_T$ similar to that in $p_T$ is observed in all regions except for the angular region R1 $\theta$=0-2°, where multi-jet events have a higher increment for $m_T$ which may be due to the elastic scattering events in *pp*-interactions. The region R1 may also be affected by the leading particle effect [12-14] with particles containing high longitudinal and low transverse momentum, as was observed in our previous study in the case of $N_{ch}$ distribution for the same polar angle region and also in the low $N_{ch}$ region for multiplicity distributions for full phase space [6]. We can see that the slopes of the distributions depend on the polar angle of the particles and the number of jets for the high $p_T$ particles ($p_T$>2 GeV/c).

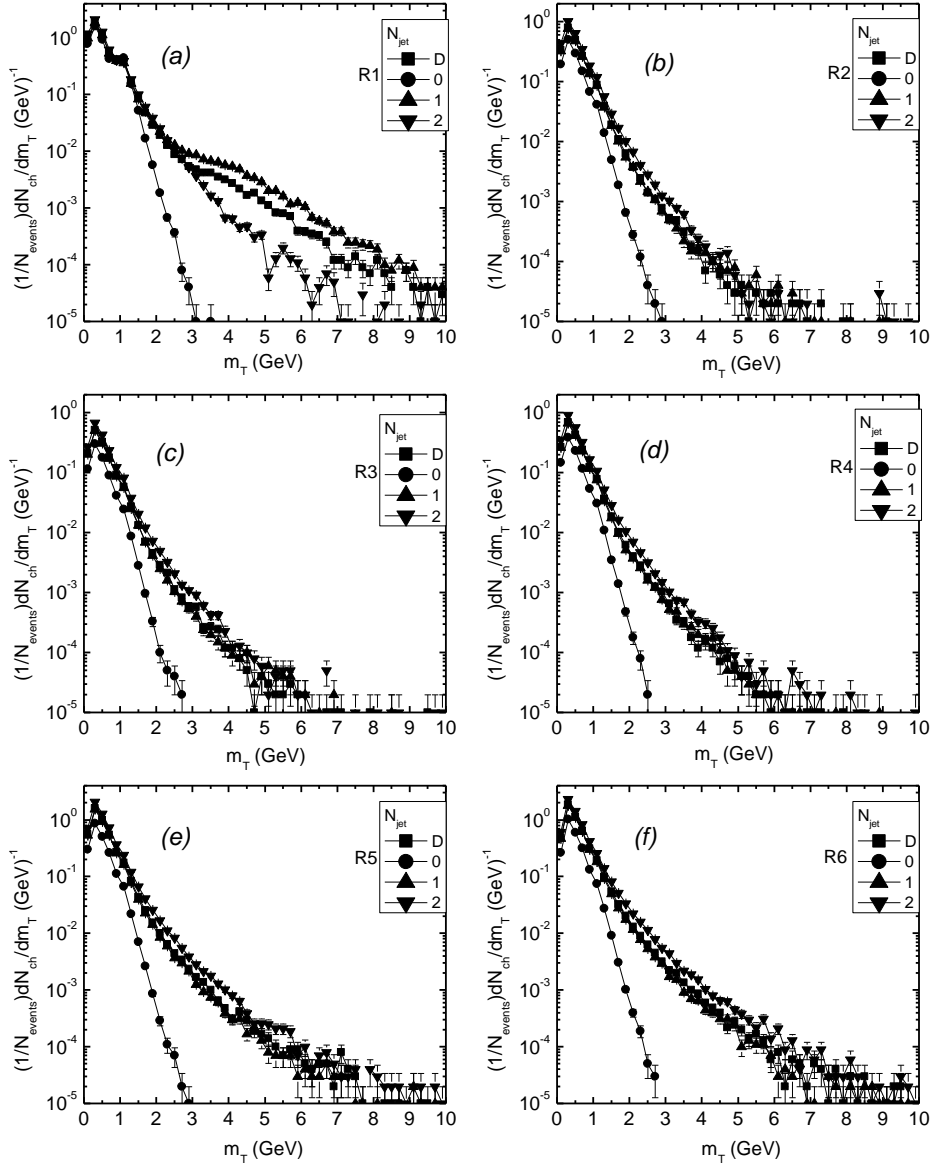

Fig. 8. $m_T$ distributions of charged particles for $N_{jet}$=D, 0, 1 and 2 in the angular regions: *(a)* R1 θ=0-2° *(b)* R2 θ=2-4° *(c)* R3 θ=4-6° *(d)* R4 θ=6-10° *(e)* R5 θ=10-30° *(f)* R6 θ=30-90°.

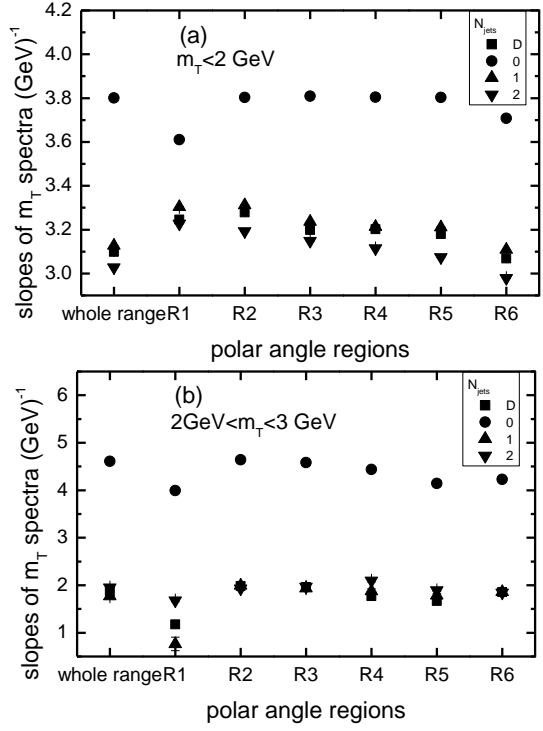

Fig. 9. Slopes of the $m_T$ spectra for the whole range and selected regions R1-R6 of the polar angle, (a) for $m_T<2$ GeV and (b) for $2$ GeV$<m_T<3$ GeV

Figure 9 ((a) and (b)) shows the plots of slopes of the $m_T$ spectra for the whole range and selected six regions R1-R6 of polar angle for $m_T<2$ GeV and for $2$ GeV$<m_T<3$ GeV as a function of number of jets. Very similar behavior to Fig. 6 is seen here. Event selection with $N_{jet}=0$ affects the $m_T$ distribution only up to $m_T \sim 3$ GeV, as for $m_T>3$ GeV events with $N_{jet}>0$ are contributing.

## 3  Conclusion

We studied the effects of jet production on the pseudorapidity, transverse momentum and transverse mass distributions of secondary charged particles produced in *pp*-collisions at 1.8 TeV in the CDF experiment, using the HIJING code. These distributions were analyzed for the whole range and for selected six regions of the polar angle as a function of different number of jets.

Some plateaus are observed in the central area of pseudorapidity. The existence of these plateaus is very important for the applicability of hydrodynamics. With increasing the number of jets the widths of the distributions decreased and the pseudorapidity density in the central area increased.

It was observed that with increasing the number of jets from 0 to 1, 2 or the default value of jets, the transverse momentum increased as the jets or multi-jet events contain high $p_T$ particles.

The behavior of the $p_T$ - distributions in the case of $N_{jet} \geq 1$ is different in two areas of $p_T$: $p_T<2$ GeV/c and $p_T>2$ GeV/c. The slopes of the distribution in the first area are very close to that for events with $N_{jet}=0$. We can therefore say that particles with $p_T<2$ GeV/c are produced by the same dynamics in events with different $N_{jet}$. The particles with $p_T>2$ GeV/c are those produced by some special dynamics, different from the particles with $p_T<2$ GeV/c. We may conclude that along with the multi-jet events, zero-jet events affect the $p_T$ spectra only up to 3 GeV/c and for $p_T>3$ GeV/c only the multi-jet events contribute. This is caused by the jet dynamics (production and hadronization of the jets) from the HIJING code. We also observed, in a previous study, similar results for the multiplicity distribution of charged particles. We concluded that the high multiplicity regions in the $N_{ch}$ distribution correspond to multi-jet events.

**References**

1  A. Majumder and M. van Leeuwen, Prog. Part. Nucl. Phys., 2011, **66**: 41-92
2  M. Spousta, Mod. Phys. Lett. A, 2013, **28**: 1330017
3  J. D. Bjorken, FERMILAB-PUB-82-059-THY, 1982
4  X-. N. Wang and M. Gyulassy, Phys. Rev. D, 1991, **44**: 3501, http://www-nsdth.lbl.gov/~xnwang/hijing/
5  M. Gyulassy and X-. N. Wang, Comp. Phys. Commun., 1994, **83**: 307
6  Ali Zaman, Mais Suleymanov, Muhammad Ajaz and Kamal Hussain Khan, Int. J. Mod. Phys. E, 2014, **23**: 1450029
7  R. Ugoccioni and A. Giovannini, J. Phys.: Conf. Ser., 2005, **5**: 199-208
8  CDF Collaboration (F. Abe et al.), Phys. Rev. D, 1990, **41**: 2330(R)
9  V. V. Uzhinsky, arXiv:hep-ph/0312089v2, http://hepweb.jinr.ru/hijing_0_1/
10  J. D. Bjorken, Phys. Rev. D, 1983, **27**: 140-151
11  A. Mueller, in Proceedings of the 1981 ISABELLE Summer Workshop, edited by H. Gordon, BNL, Upton, New York, 1982, 636
12  M. Ajaz, M. K. Suleymanov, K. H. Khan, A. Zaman, H. Younis and A. Rahman, Mod. Phys. Lett. A, 2013, **28**: 1350175
13  M. Ajaz, M. K. Suleymanov, K. H. Khan and A. Zaman, Int. J. Mod. Phys. E, 2012, **21**: 1250095.
14  T. Mizoguchi, T. Aoki, M. Biyajima and N. Suzuki, Prog. Theor. Phys., 1992, **88**: 391